\pdfoutput=1
\documentclass[sigconf,screen=true,bookmarks=false,9pt]{acmart}
\usepackage{graphicx}
\usepackage{amsmath}

\usepackage{amssymb}
\usepackage{mathtools}
\usepackage{comment}
\usepackage[subrefformat=parens,labelformat=parens]{subfig}
\usepackage{bm}
\usepackage{multirow}
\usepackage{threeparttable,booktabs}
\usepackage{tikz}
\usepackage{balance}
\usepackage{courier}                                       % courier font, used in \texttt
\usepackage{cleveref}                                      % smart citation
\usepackage[mathcal]{eucal}
\usepackage[noend]{algpseudocode}
\algrenewcommand\textproc{\texttt}
\makeatletter\let\float@addtolists\relax\makeatother
\usepackage{algorithm}
\usepackage{filecontents}                                  % support to pgfplots
\usepackage{pgfplots}
\usepackage{pgfplotstable}
\pgfplotsset{compat=newest}
\usepackage{siunitx}

\renewcommand{\vec}[1]{\boldsymbol{#1}}

\theoremstyle{plain}

\theoremstyle{definition}

\usepackage[skip=1pt]{caption}            % set space between figure and caption
\graphicspath{{./figs/}}

\usepackage{color}
\usepackage{xcolor}
\definecolor{ACMPuple}{cmyk}{0.55,1,0,0.15}

\newcommand\blfootnote[1]{%
  \begingroup
  \renewcommand\thefootnote{}\footnote{#1}%
  \addtocounter{footnote}{-1}%
  \endgroup
}

\copyrightyear{2024}
\acmYear{2024}
\setcopyright{acmlicensed}\acmConference[DAC '24]{61st ACM/IEEE Design
Automation Conference}{June 23--27, 2024}{San Francisco, CA, USA}
\acmBooktitle{61st ACM/IEEE Design Automation Conference (DAC '24), June
23--27, 2024, San Francisco, CA, USA}
\acmDOI{10.1145/3649329.3656229}
\acmISBN{979-8-4007-0601-1/24/06}

\begin{document}

\settopmatter{printacmref=false} % Removes citation information below abstract
\pagestyle{plain} % removes running headers

%\renewcommand{\baselinestretch}{1}

% \copyrightyear{2019} 
% \acmYear{2019} 
% \setcopyright{acmcopyright}
% \acmConference[DAC '19]{The 56th Annual Design Automation Conference 2019}{June 2--6, 2019}{Las Vegas, NV, USA}
% \acmBooktitle{The 56th Annual Design Automation Conference 2019 (DAC '19), June 2--6, 2019, Las Vegas, NV, USA}
% \acmPrice{15.00}
% \acmDOI{10.1145/3316781.3317803}
% \acmISBN{978-1-4503-6725-7/19/06}

\title{
\textbf{
%Deep Learning Toolkit Enabled Placement Acceleration on GPU with Massive Parallelism
EasyACIM: An End-to-End Automated Analog CIM with Synthesizable Architecture and Agile Design Space Exploration
%with \underline{M}assive Parallelism
}}

\author{
    Haoyi Zhang$^1$,
    Jiahao Song$^1$, 
    Xiaohan Gao$^3$, \\
    Xiyuan Tang$^{1,2}$, 
    Yibo Lin$^{1,4,5*}$, 
    Runsheng Wang$^{1,4,5}$, 
    Ru Huang$^{1,4,5}$ \\
    $^1$School of Integrated Circuits
    $^2$Institute for Artificial Intelligence, Peking University \\ 
    $^3$School of Computer Science 
    $^4$Beijing Advanced Innovation Center for Integrated Circuits  \\
    $^5$Institute of Electronic Design Automation, Peking University, Wuxi, China \\
    % $^3$Beijing Advanced Innovation Center for Integrated Circuits, Beijing, China \\
    { hy.zhang@stu.pku.edu.cn, yibolin@pku.edu.cn}
}

% \iftrue
% \author{Yibo Lin}
% \affiliation{
%     \institution{ECE Department, UT Austin}
% }
% \email{yibolin@utexas.edu}

\begin{abstract}
    Analog Computing-in-Memory~(ACIM) is an emerging architecture to perform efficient AI edge computing. However, current ACIM designs usually have unscalable topology and still heavily rely on manual efforts. These drawbacks limit the ACIM application scenarios and lead to an undesired time-to-market. This work proposes an end-to-end automated ACIM based on a synthesizable architecture (EasyACIM). With a given array size and customized cell library, EasyACIM can generate layouts for ACIMs with various design specifications end-to-end automatically. Leveraging the multi-objective genetic algorithm (MOGA)-based design space explorer, EasyACIM can obtain high-quality ACIM solutions based on the proposed synthesizable architecture, targeting versatile application scenarios. The ACIM solutions given by EasyACIM have a wide design space and competitive performance compared to the state-of-the-art (SOTA) ACIMs.
\end{abstract}

% \keywords{Analog Computing-in-Memory, Synthesizable Architecture, }
\maketitle
\blfootnote{$^*$Corresponding author}

\vspace{-0.8cm}
\section{Introduction}
\label{sec:Introduction}

% placement is important 
% what is placement 
% state-of-the-art placement 
% previous parallelization effort 
% our contribution 
\begin{figure}[tb]
    \centering
    \includegraphics[width=0.35\textwidth]{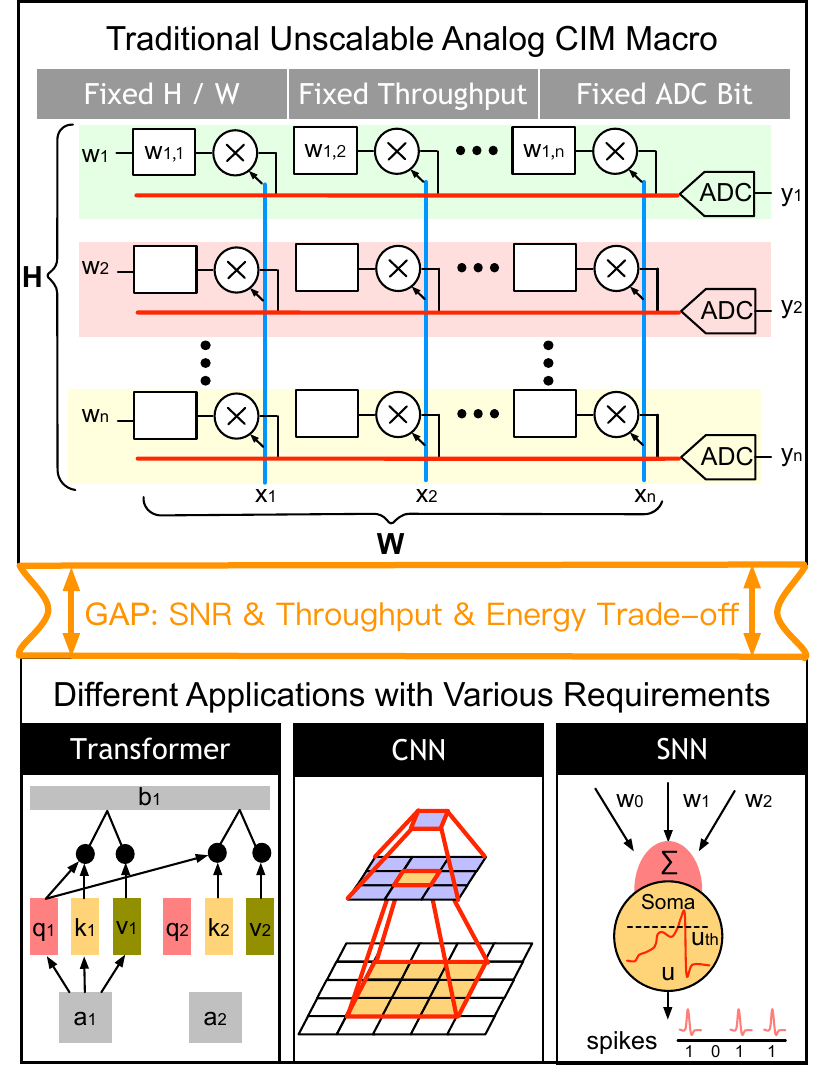}
    \caption{Unscalable ACIM macro and various scenarios.}
    \label{Intro}
    \vspace{-0.7cm}
\end{figure}

\begin{sloppypar}
With the emergence of AI technology, the demand for computility has increased dramatically and the memory-wall effect is becoming more and more evident. Computing-in-Memory~(CIM) is a popular solution for AI accelerators addressing the memory bottleneck. Exploiting the structural alignment between a dense 2D array of bit cells and the dataflow in matrix-vector multiplication, CIM has unique advantages in energy and throughput over other solutions~\cite{vermaInMemoryComputingAdvances2019}. The mainstream CIM can be categorized into two groups ACIM and Digital CIM~(DCIM). Although DCIM has better robustness, ACIM still has great potential because of high energy efficiency and high density at lower computing precision~\cite{jhangChallengesTrendsSRAMBased2021}. Based on this unique feature, ACIM is able to occupy a niche in AI edge computing. 
\end{sloppypar}

\begin{sloppypar}
Traditional ACIM research has focused on the pursuit of extreme performance such as high energy efficiency~\cite{cheon2941TOPSChargeDomain10T2023,yaoFullyBitFlexibleComputation2023a}, high area efficiency~\cite{yu65nm8TSRAM2022,zhangHelloEdgeKeyword2018a}, high accuracy~\cite{yaoFullyBitFlexibleComputation2023a,yan041MbMm272022} or high throughput~\cite{dong15351TOPS3722020}. As Figure~\ref{Intro} shows, these designs often have an unscalable topology with fixed array height $H$, array width $W$, and ADC bits $B_\text{ADC}$. These fixed parameters lead to the gap between the unscalable CIM macro and different application scenarios. For example, a transformer for large language model~(LLM) and a convolution neural network~(CNN) for image identification are likely to have different accuracy requirements. A particular CIM macro may not be accurate enough for the transformer but has energy waste for CNN due to the redundant accuracy. While some works have flexible ADC bits~\cite{yu65nm8TSRAM2022,yuLogicCompatibleEDRAMComputeInMemory2021,ali65Nm462023}, it is difficult for such reconfigurable designs to eliminate all the overhead caused by redundant precision, including area, energy consumption, and throughput.
\end{sloppypar}

% A typical design flow for ACIM is to first rely on human experts to determine various design parameters, and then design the circuits and layouts with these parameters fixed.

\begin{sloppypar}
Beyond the ACIM circuit design itself, the design efficiency is also very significant. As the electronic design automation (EDA) technology evolves, some studies have emerged to help CIM circuits benchmarking~\cite{chenNeuroSimCircuitLevelMacro2018,sunAnalogDigitalInmemory} and modeling~\cite{gonugondlaFundamentalLimitsPrecision2020,gonugondlaFundamentalLimitsEnergyDelayAccuracy2022}. Furthermore, inspired by end-to-end automated flow for SRAM~\cite{kamineniMemGenOpenSourceFramework2021} and SAR ADC~\cite{liuOpenSAROpenSource2021}, AutoDCIM~\cite{chenAutoDCIMAutomatedDigital}
proposed the first end-to-end automated flow for DCIM. However, the end-to-end automated flow for ACIM is still a blank slate since ACIM has a more sophisticated signal-noise ratio (SNR), energy, area, and throughput trade-off strategies than DCIM. Therefore, a complete end-to-end flow for ACIM should automatically optimize these trade-offs rather than leaving them up to users, as is the case with AutoDCIM~\cite{chenAutoDCIMAutomatedDigital}.
\end{sloppypar}

\begin{sloppypar}
In this work, we propose EasyACIM, an end-to-end automated ACIM with a fully synthesizable ACIM architecture and agile exploration of design specifications. To narrow the gap between the CIM macro and different application scenarios, EasyACIM proposes a novel ACIM architecture that can easily be implemented with different $H$, $W$, $B_\text{ADC}$, and throughput. EasyACIM constructs an estimation model for the particular architecture and leverages the genetic algorithm to automatically explore the Pareto frontier for the synthesizable architecture with a given array size. Such an approach further improves the design efficiency which is ignored in AutoDCIM~\cite{chenAutoDCIMAutomatedDigital}. EasyACIM integrates a template-based hierarchical placement and routing framework to generate the final layouts for the ACIM with an agile exploration of design specifications. The main contributions of this paper can be summarized as follows: 
\begin{itemize}
    \item  We propose a novel synthesizable ACIM architecture leveraging the local compute array and the reusable capacitors that can be used as CDAC capacitors in SAR ADCs, which is easily implemented into versatile application scenarios. 
    \item We treat the determination of ACIM parameters as a multi-objective optimization problem, build estimation models for the proposed ACIM, and obtain the Pareto frontier by a MOGA-based (NGSA-II) design space explorer. 
    \item We integrate a template-based hierarchical placement and routing framework into the EasyACIM, in order to generate the final layouts according to the Pareto-frontier design specifications. 
    \item As illustrated in the results, EasyACIM can generate ACIMs with SOTA performance for various applications with a wide design space where the energy efficiency ranges from 50TOPS/W to 750TOPS/W and the area ranges from 1500$\text{F}^2$/bit to 7500$\text{F}^2$/bit. 

\end{itemize}
\end{sloppypar}
% The source code is released on Github
% To clarify, %this work focuses on accelerating analytical placement leveraging deep learning toolkit and GPUs, which is a orthogonal to using deep learning for placement. 
% the casting of placement problem to deep learning problems aims at using the toolkit to solve placement, which is orthogonal to using deep learning models for placement. 
%It is irrelevant to any application of deep learning algorithms. 
\begin{sloppypar}
The rest of the paper is organized as follows. 
Section~\ref{sec:Preliminary} describes the background; 
Section~\ref{sec:Algorithm} explains the detailed implementation; 
Section~\ref{sec:Results} demonstrates the results; 
Section~\ref{sec:Conclusion} concludes the paper. 
\end{sloppypar}

%\textcolor{blue}{Link each technique to the speedup achieved. }
%\vspace{-.05in}
\section{Preliminaries}
\label{sec:Preliminary}

This section will review the background for the ACIM compute model, Pareto optimization, and layout automation for Analog and Mixed Signal (AMS) design, respectively. 

\subsection{ACIM Compute Model}
\begin{figure}[tb]
    \centering
    \includegraphics[width=0.3\textwidth]{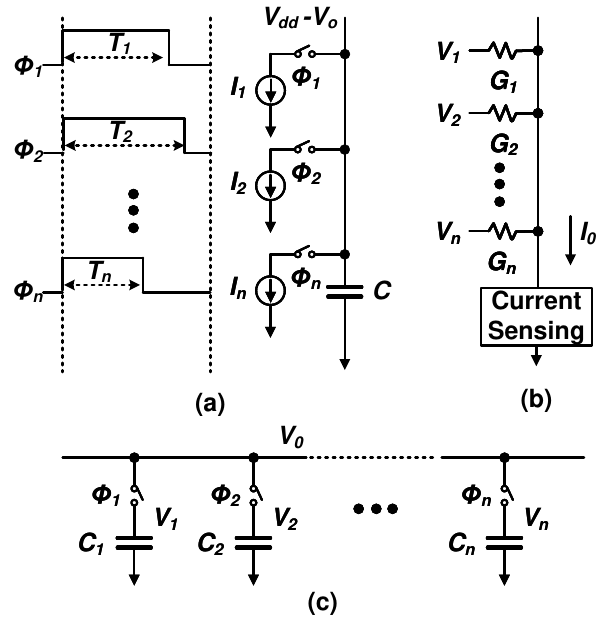}
    \caption{In-memory compute models: (a) QS (b) IS (c) QR.}
    \label{ACIMType}
    \vspace{-0.3cm}
\end{figure}

Much research on ACIM has emerged in recent years. Most of them employ following three in-memory compute models (Figure~\ref{ACIMType}): (a) charge summing~(QS)~\cite{kangMultiFunctionalInMemoryInference2018}; (b) current summing~(IS)~\cite{yinXNORSRAMInMemoryComputing2020}; (c) charge redistribution~(QR)~\cite{rastegariXNORNetImageNetClassification2016}. QS and QR are both charge-domain compute models that are insensitive to the process-voltage-temperature~(PVT). However, such an approach stores information in the form of electrical charges and requires additional metal capacitance, resulting in additional area overhead. In more detail, the QR model leverages the redistributing charge between storage units which is more flexible and extensible for different computing applications. The QS model generates the results by summing the charge from the storage units which is more difficult to support different applications. IS is a current-domain compute model that usually has higher density but is sensitive to PVT. The information is stored in the electric current which is also difficult to be adaptive with various applications. For the consideration of robustness and extensibility, EasyACIM selects QR as the compute model for synthesizable architecture.

\subsection{Pareto Optimization}

The trade-off among SNR, energy, throughput, and area in ACIM is a typical muti-objective optimization problem~\cite{pereiraReviewMultiobjectiveOptimization2022}. It is difficult to find a single optimal solution, especially when faced with different application scenarios. Therefore, obtaining the Pareto-frontier set for the ACIM is a feasible solution. The Pareto-frontier set is made up of the solution vectors that are not dominated by other vectors. Formally, a solution vector $\vec{u}=\left[u_{1}, u_{2}, \ldots, u_{P}\right]^{T}$ is said to pareto-dominate~\cite{ngatchouParetoMultiObjective2005} the solution vector $\vec{v}=\left[v_{1}, v_{2}, \ldots, v_{P}\right]^{T}$, in a minimization context, if and only if: 

\vspace{-0.2cm}
\begin{equation}
    \begin{array}{l}
        \forall i \in\{1, \ldots, N\}, f_{i}(\vec{u}) \leq f_{i}(\vec{v}) \\
        \text { and } \exists j \in\{1, \ldots, N\}: f_{j}(\vec{u})<f_{j}(\vec{v})
        \end{array}
    \label{ParetoFront}
\end{equation}

The function values of the Pareto-frontier set form the Pareto frontier for a particular multi-objective optimization problem.  

% \begin{figure}[tb]
%     \centering
%     \includegraphics[width=0.25\textwidth]{figs/Fig3.ParetoFront.pdf}
%     \caption{A Pareto frontier example.}
%     \label{ParetoFront}
%     \vspace{-0.4cm}
% \end{figure}

\subsection{Layout Automation for AMS Design}
The layout design of ACIM is more similar to analog and mixed-signal (AMS) circuits than digital circuits since the ACIM includes versatile AMS blocks such as SAR ADC, sense amplifier (SA), and CMOS switch. Such AMS blocks prevent the designers from using commercial digital layout automation tools to generate ACIM layouts. Therefore, the layout automation tools for AMS circuits are more suitable when tackling ACIM layouts.

\begin{figure}[h]
    \centering
    \includegraphics[width=0.4\textwidth]{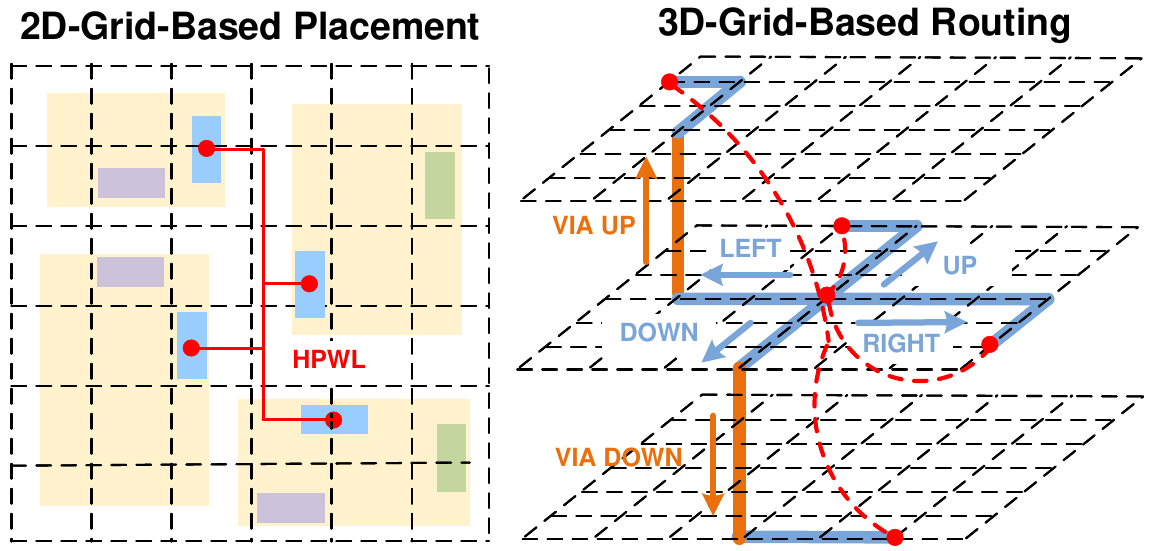}
    \caption{Basic grid-based placement and routing.}
    \label{GridPR}
    \vspace{-0.3cm}
\end{figure}

Much research has been done on the placement and routing problems of AMS circuit designs. ALIGN~\cite{kunalINVITEDALIGNOpenSourcea} and MAGICAL~\cite{chenMAGICALOpenSource2021} both construct a complete framework for AMS designs including placement and routing. These frameworks already have the ability to generate decent layouts for AMS designs. After that some independent placement and routing targeting better performance have emerged, such as SAGERoute~\cite{zhangSAGERouteSynergisticAnalog,zhangSAGERoute2.0SynergisticAnalog} and hierechical AMS placement~\cite{zhuHierarchicalAnalogMixedSignal2023}. All of these placement and routing methodologies are based on the partitioned grids, as Figure~\ref{GridPR} shows. Since the grid-based method is easier to extend with versatile scenarios and honors different constraints in AMS design layouts. In practice, more constraints such as symmetry, alignment, etc. should be considered in addition to the basic half-perimeter wire length (HPWL). 

Although academia is booming in AMS layout automation, the automatically-generated layouts may still be unsatisfactory in some extreme cases. For example, the SRAM cell in ACIM is very dense, and the routing track is often well-designed by experienced designers. It is difficult for a fully automated tool to meet all the requirements. Therefore, we develop a template-based placement and routing method along with automated approaches to generate high-quality layout solutions for ACIM utilizing manually designed layout cells.
\section{EasyACIM Framework}
\label{sec:Algorithm}

\begin{figure}[b]
    \vspace{-0.2cm}
    \centering
    \includegraphics[width=0.33\textwidth]{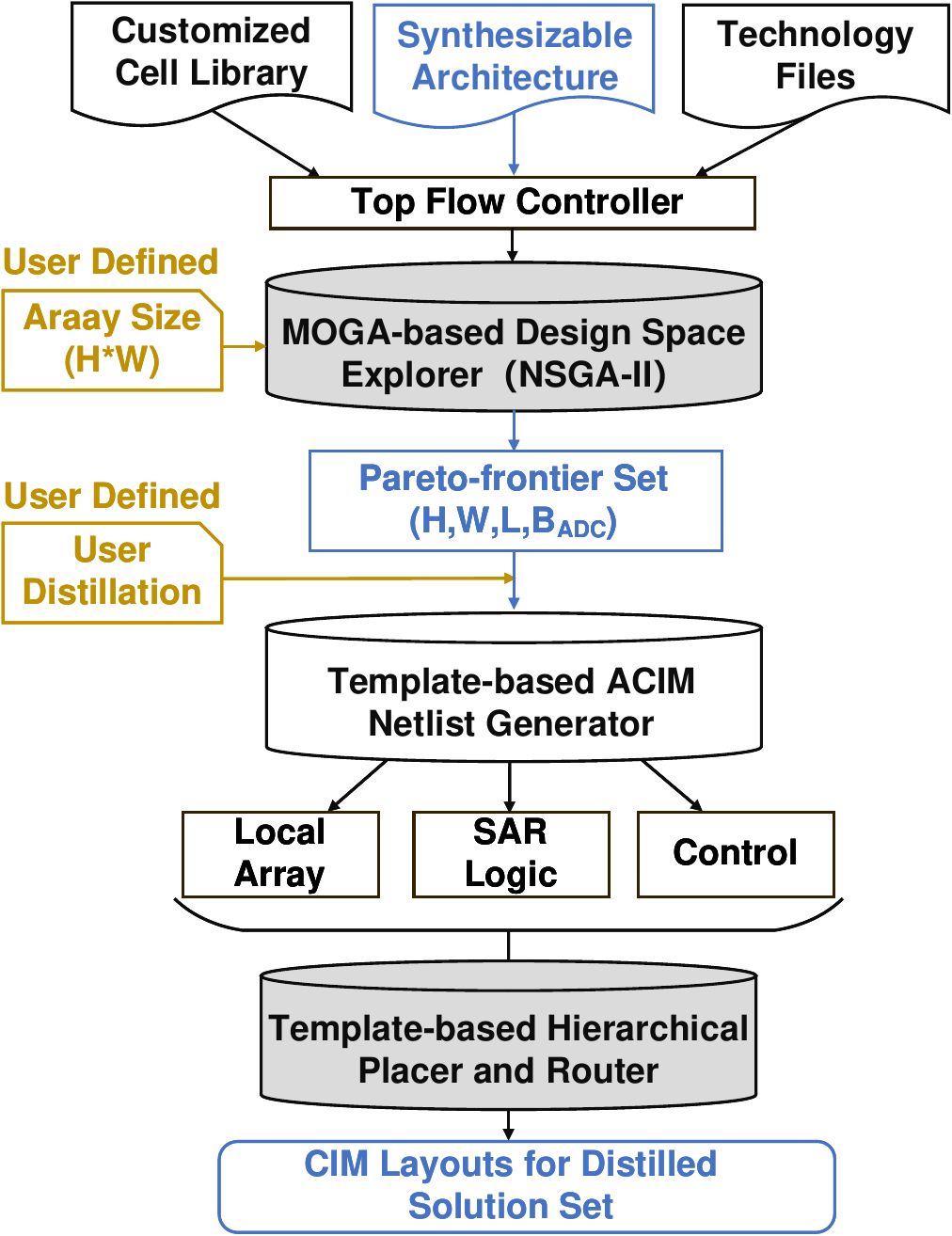}
    \caption{Overview of EasyACIM.}
    \label{FlowFramework}
    \vspace{-0.7cm}
\end{figure}

\begin{sloppypar}
An overview of the EasyACIM framework is depicted in Figure~\ref{FlowFramework}. The whole framework takes a customized cell library, synthesizable architecture, and technology files as input. The customized cell library includes netlists of all the components of ACIM (e.g. SAR logic, SA, 8T SRAM cell) and layouts of critical components of ACIM (e.g. SA, 8T SRAM cell). The synthesizable architecture determines the rules for combining these components. The technology files contain the necessary information for layout generation (e.g. DRC rules, layer map). 
\end{sloppypar}

\begin{sloppypar}
The MOGA-based design space explorer can generate a Pareto-frontier set at a user-defined array size leveraging NSGA-II, a classic MOGA. Each solution vector in the Pareto-frontier set contains four components including array height ($H$), array width ($W$), local array size ($L$), and ADC precision bits ($B_{\text {ADC}}$). After the automatic exploration, the users can remove undesired solutions from the Pareto-frontier set according to their requirements. Via this agile interaction, the Pareto-frontier set can be further refined to match the desired application scenarios. Then the template-based netlist generator as well as template-based hierarchical placer and router will be conducted in sequence for each solution of the Pareto-frontier set. Finally, high-quality ACIM layouts can be generated, ensuring Pareto-frontier design specifications that align with the user's requirements.
\end{sloppypar}

\subsection{Synthesizable Architecture Design}
\begin{figure*}[tb]
    \centering
    \includegraphics[width=0.9\textwidth]{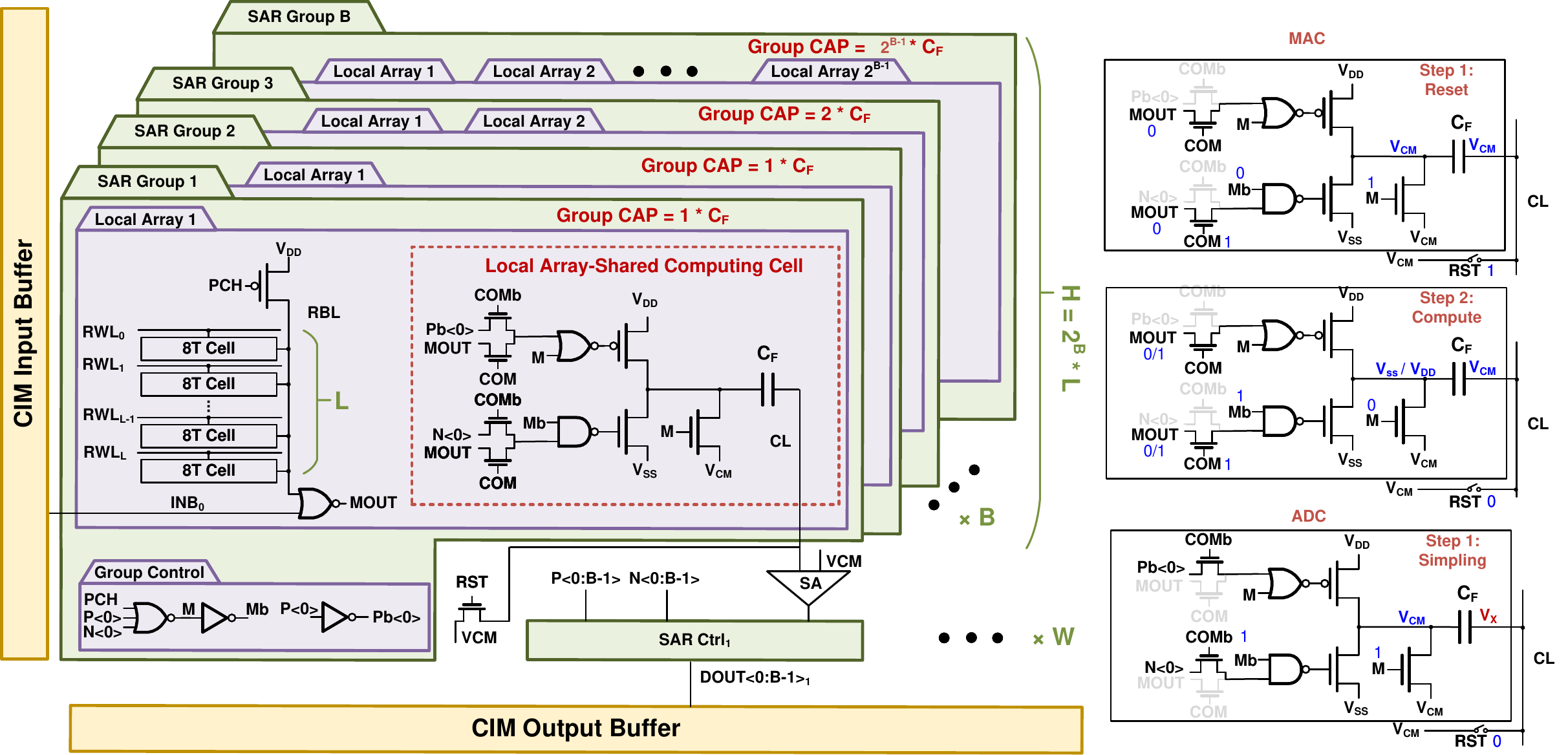}
    \caption{The synthesizable architecture and operating states.}
    \label{CIMArchitecture}
    \vspace{-0.3cm}
\end{figure*}

\begin{sloppypar}
Figure~\ref{CIMArchitecture} demonstrates the overview of the proposed synthesizable ACIM architecture as well as the basic operating states. One column of the proposed ACIM is detailed in Figure~\ref{CIMArchitecture}. Inspired by a novel design~\cite{yaoFullyBitFlexibleComputation2023a}, we reuse the compute capacitors~$C_\text{F}$ as the capacitors in CDAC during the SAR ADC conversion. This is achieved by dividing them into distinct SAR groups following a ratio of 1:1:2:4:$\cdots$:$2^n$, aligning with the capacitance ratio in the CDAC. This approach greatly reduces the ADC area overhead in the ACIM designs. However, if each 8T SRAM cell is furnished with an individual capacitor and its corresponding control circuit, the area overhead will remain substantial. Therefore, we combine $L$ 8T-SRAM cells into a local array~\cite{rastegariXNORNetImageNetClassification2016}. The $L$ 8T-SRAM cells in a particular local array share the same compute capacitor and control circuits. Selecting an appropriate $L$ introduces a trade-off between area and throughput.  
\end{sloppypar}

\begin{figure}[tb]
    \centering
    \includegraphics[width=0.33\textwidth]{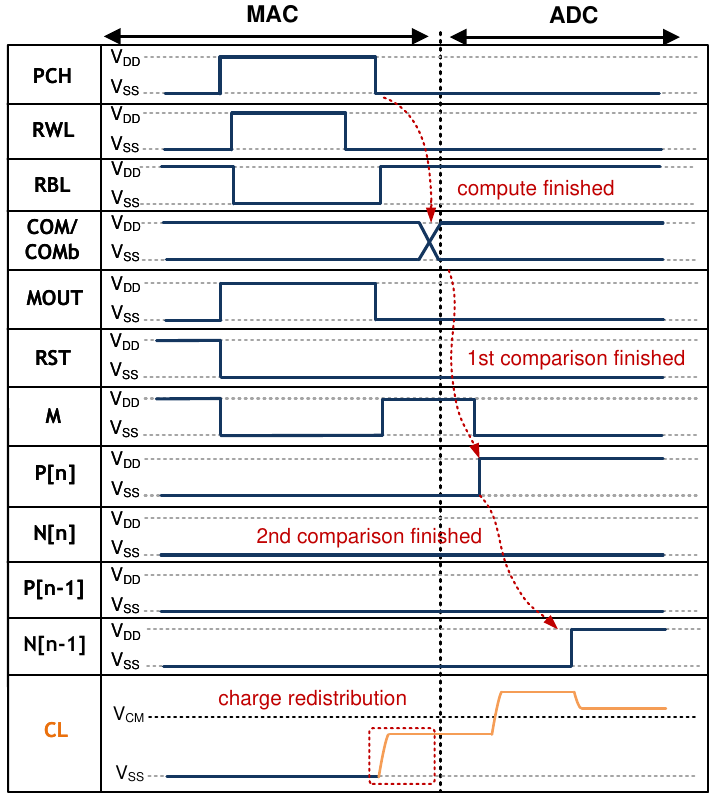}
    \caption{Timing diagram of the synthesizable ACIM.}
    \label{CIMTimeflow}
    \vspace{-0.7cm}
\end{figure}

\begin{sloppypar}
The proposed ACIM architecture has two operating states: 1) multiply-accumulate (MAC) state, and 2) ADC conversion state. In the MAC state, both ends of the capacitor $C_\text{F}$ are reset to the $V_\text{CM}$ at first. Then, as Figure~\ref{CIMTimeflow} shows, the RWL turns to $V_\text{dd}$, the RST turns to $V_\text{ss}$, and the MAC operation starts. After the MAC operation, the top plate of the capacitor will be changed to either $V_\text{dd}$ or $V_\text{ss}$, representing the computation result. In the ADC conversion state, the top plate will be reset to $V_\text{CM}$ again and the charge will redistribute in the bottom plate of the capacitor. After charge redistribution, the final accumulation result $V_\text{x}$ is stored on the RBL. Then the SAR logic starts the switching procedure to get the final digital result. The P[n] and N[n] is the switching control signal based on the comparison results of each round. After $B_{\text {ADC}}$ rounds comparison, the final MAC result with the precision of $B_{\text {ADC}}$ bits can be obtained. When dealing with a different $B_{\text {ADC}}$, a CMOS switch will be inserted in the appropriate position of the RBL. The CMOS switch will remain closed until the charge redistribution is complete. Then the CMOS switch will be opened to separate the redundant large capacitance, thus saving energy during the ADC conversion process. 
\end{sloppypar}

\subsection{MOGA-based Design Space Explorer}
\begin{sloppypar}
The MOGA-based design space explorer consists of two parts: 1) ACIM performance estimation model and 2) NSGA-II-based optimization. While exploring ACIM design specifications, constructing the ACIM performance estimation model is more important than the optimization algorithm itself. The primary focus lies in developing an accurate and efficient ACIM estimation model, which is our main concentration. With a robust ACIM estimation model, the requirements of the algorithm can be appropriately relaxed. We select NSGA-II as the exploration algorithm due to its adeptness in maintaining a superior balance, facilitating smooth convergence, and preserving diversity within solutions~\cite{pereiraReviewMultiobjectiveOptimization2022}.
\end{sloppypar}

\subsubsection{ACIM Performance Estimation Model}
\renewcommand{\arraystretch}{1.5}
\begin{sloppypar}
The ACIM can be evaluated from various perspectives, such as SNR, energy, area, and throughput. EasyACIM adopts a QR mode computation with bottom-plate charge redistribution. Reference to literature~\cite{gonugondlaFundamentalLimitsEnergyDelayAccuracy2022}, a customized estimation model is constructed for EasyACIM and is detailed as follows.
\end{sloppypar}

The total SNR ($\mathrm{SNR}_{\mathrm{T}}$) is shown in Equation~\ref{equ:SNRTotal}, where $\mathrm{SNR}_{\mathrm{pre}}$ indicates the SNR before ADC, and $\mathrm{SQNR}_{\mathrm{y}}$ indicates the SNR for quantization noise of ADC. 
\begin{equation}
    \label{equ:SNRTotal}
    \mathrm{SNR}_{\mathrm{T}}=\left[\frac{1}{\mathrm{SNR}_{\mathrm{pre}}}+\frac{1}{\mathrm{SQNR}_\mathrm{y}}\right]^{-1}
  \end{equation}

The $\mathrm{SNR}_{\mathrm{pre}}$ can be breakdown into $\mathrm{SNR}_{\mathrm{a}}$ and $\mathrm{SQNR}_{\mathrm{i}}$. The $\mathrm{SNR}_{\mathrm{a}}$ is the noise caused by the analog circuits and $\mathrm{SQNR}_{\mathrm{i}}$ is the output referred SQNR due to input (weight and activation) quantization. 
\begin{equation}
    \label{equ:SNRPre}
    \mathrm{SNR}_{\mathrm{pre}}=\frac{\sigma_{y_{\mathrm{o}}}^{2}}{\sigma_{q_{i}}^{2}+\sigma_{\eta_{\mathrm{a}}}^{2}}=\left[\frac{1}{\mathrm{SNR}_{\mathrm{a}}}+\frac{1}{\mathrm{SQNR}_\mathrm{i}}\right]^{-1}
\end{equation}

In equation~\ref{equ:SNRPre}, $\sigma_{y_{\mathrm{o}}}^{2}$ is the variance of the output and $\sigma_{y_{\mathrm{o}}}^{2}=N \sigma_{w}^{2} \mathbb{E}\left[x^{2}\right]$. $\sigma_{q_{i}}^{2}$ is the variance of input quantization noise, $\sigma_{\eta_{\mathrm{e}}}^{2}$ is the variance of analog nonlinearity. Their definitions are detailed in \ref{equ:qi} and \ref{equ:ne}, respectively. Table~\ref{tab:notation} gives definitions of some basic symbols.

\renewcommand{\arraystretch}{0.9}
\begin{table}[b]
    \vspace{-0.4cm}
    \centering
    \caption{\textsc{Notation}}
    \resizebox{0.4\textwidth}{!}{
    \begin{tabular}{cc}
    \toprule
    Symbol    &  Description\\
    \midrule
    $N$ & dot product length \\ 
    $B$ & precision in bits \\ 
    $x,w,$ and $y$ & inputs, weights, and outputs \\
    $x_{\mathrm{m}}, w_{\mathrm{m}},$ and $y_{\mathrm{m}}$ & the corresponding maximum \\
    $\sigma_{x}, \sigma_{w}$ & standard deviation of input and weight \\ 
    \bottomrule
    \end{tabular}
    }
    \label{tab:notation}
\end{table}

\renewcommand{\arraystretch}{1.2}

\begin{equation}
    \label{equ:qi}
    \sigma_{q_{i}}^{2}=\frac{1}{12} N \Delta_{x}^{2} \sigma_{w}^{2}+\frac{1}{12} N \Delta_{w}^{2} \mathbb{E}\left[x^{2}\right] 
\end{equation}

In Equation~\ref{equ:qi} $\Delta_{w}=w_{\mathrm{m}} 2^{-B_{w}+1}  ,  \;  \Delta_{x}=x_{\mathrm{m}} 2^{-B_{x}}$ 

\begin{equation}
    \label{equ:ne}
    \sigma_{\eta_{\mathrm{e}}}^{2}=\frac{2}{3}\left(1-4^{-B_{w}}\right) N\left(\frac{\mathbb{E}\left[x^{2}\right] \sigma_{C_{0}}^{2}}{C_{\mathrm{o}}^{2}}+\frac{2 \sigma_{\theta,\mathrm{o}}^{2}}{V_{\mathrm{dd}}^{2}}+\sigma_{\mathrm{inj}}^{2}\right)
\end{equation}

In Equation~\ref{equ:ne}, $C_{\mathrm{o}}$ is the compute capacitor with standard deviation $\sigma_{C_{\mathrm{o}}}=\kappa \sqrt{C_{\mathrm{o}}}$. $\kappa$ is a layout and technology dependent mismatch coefficient~\cite{tripathiMismatchCharacterizationSmall2014}. $\sigma_{\theta, \mathrm{o}}=\sqrt{\frac{k T}{C_{\mathrm{o}}}}$ is the themeral noise caused by $C_{\mathrm{o}}$, $k$ is the Boltzmann constant, $T$ is the temperature in Kelvin. $\sigma_{\mathrm{inj}}^{2}$ indicates the noise caused by charge injection which is almost eliminated by the bottom-plate charge redistribution technique and can be ignored in the following calculations. $\text{SQNR}_\text{y}$ is detailed as follows, where $\zeta_{x}=x_{m} / \sigma_{x} ,  \;  \zeta_{w}=w_{m} / \sigma_{w}$. 
% \begin{equation}
%     \label{equ:yo}
%     \sigma_{y_{\mathrm{o}}}^{2}=N \sigma_{w}^{2} \mathbb{E}\left[x^{2}\right]
% \end{equation}

\begin{equation}
    \label{equ:SQNR}
    \begin{array}{l}\operatorname{SQNR}_{\mathrm{y}(\mathrm{dB})}=10 \log _{10}\left(\frac{\sigma_{y_{\mathrm{o}}}^{2}}{\sigma_{q_{y}}^{2}}\right) \\\quad=6 B_{y}+4.8-\left[\zeta_{x(\mathrm{dB})}+\zeta_{w(\mathrm{dB})}\right]-10 \log _{10}(N)\end{array}
\end{equation}

The throughput can be described as Equation~\ref{equ:throughput}. $t_\text{com}$ is the computation delay which is much less than ADC's delay. The delay of ADC can be broken down into setup time $t_\text{set}$ and 1-bit conversion time $t_\text{conv}$. $t_\text{set}$ should satisfy $t_{\text{set}}>0.69 \tau B_{\text{ADC}}$, where $\tau$ is the time constant. $t_{\text{conv}}$ can be estimated by $t_{\text{conv}}=t_{\text{conv/bit}} \cdot B_{\text{ADC}}$
\begin{equation}
    \label{equ:throughput}
    T=\frac{H}{L} \cdot W / (t_\text{com}+t_\text{set}+t_\text{conv}) \\
\end{equation}

The total average energy for one-bit computing can be defined as Equation~\ref{equ:energy}. The $E_{\text {compute}}$ and $E_{\text {control}}$ are almost constant for different design specifications in a given architecture. The power consumption of ADC with different precision really makes the difference.  

\begin{equation}
    \label{equ:energy}
E=E_{\text {compute}}+E_{\text {control}}+\frac{E_\text{ADC}}{H / L} \\
\end{equation}

The $E_{\mathrm{ADC}}$ has an empirical formula~\cite{EnergyADC} described as Equation~\ref{equ:EADC}, where $k_{1}$ and $k_{2}$ are empirical parameters which can be obtained from post-layout simulation.
\begin{equation}
    \label{equ:EADC}
    E_{\mathrm{ADC}}=k_{1} \cdot \left(B_{\mathrm{ADC}}+\log_{2}{V_{\mathrm{DD}}}\right)+k_{2} \cdot 4^{B_{\mathrm{ADC}}} \cdot V_{\mathrm{DD}}^{2}
\end{equation}

The average area of ACIM is demonstrated in Equation~\ref{equ:area}, where $A_{\text{SRAM}}$ is the 8T-SARM cell area, $A_{\text{LC}}$ is the area of local array-shared computing cell, $A_{\text {COMP}}$ is the area of the dynamic comparator and $A_{\text {DFF}}$ is the area of a single dynamic D-type Flip Flop~(DFF) in the SAR logic. 

\begin{equation}
    \label{equ:area}
A=A_{\text{SRAM}}+ \frac{1}{L} \cdot A_{\text{LC}}+ \frac{1}{H} \cdot A_{\text {COMP }} + \frac{1}{H} \cdot B_{\text{ADC}} \cdot A_{\text {DFF}}
\end{equation}

% \begin{array}{l}
% A=A_{S C}+A_{L C}+A_{S R A M}+A_{\text {comp }} \\
% A=A_{50^{\circ}} \cdot B_{A D C}+A_{L 0} \cdot W \cdot \frac{H}{L}
% \end{array}

% \begin{equation}
%     \sigma_{C_{j}}=\kappa \sqrt{C_{j}}, v_{j}=p \frac{W L C_{\mathrm{ox}}\left(V_{\mathrm{dd}}-V_{\mathrm{t}}-V_{j}\right)}{C_{j}}, \sigma_{\theta, j}=\sqrt{\frac{k T}{C_{j}}}
% \end{equation}

% \begin{equation}
%     \sigma_{\mathrm{inj}}^{2}=\mathbb{E}\left[x^{2}\right] W L C_{\mathrm{OX}} / C_{\mathrm{o}}
% \end{equation}

% \begin{equation}
% \begin{array}{l} 
%     D_{A D C}= t_{\text {set }}+t_{\text {conv }} \\
%     t_{\text {set }}>0.69 \tau \mathrm{ADC}_{\text {res }} \\
%     t_{\text {conv }}=t_{\text {conv/bit }} \cdot \mathrm{ADC}_{\text {res }}
% \end{array}
% \end{equation}

% \begin{equation}
%     \zeta_{x}=x_{m} / \sigma_{x} ,  \;  \zeta_{w}=w_{m} / \sigma_{w}
% \end{equation}

% \begin{equation}
%     \Delta_{w}=w_{\mathrm{m}} 2^{-B_{w}+1}  ,  \;  \Delta_{x}=x_{\mathrm{m}} 2^{-B_{x}} 
% \end{equation}

\subsubsection{NSGA-II-based optimization}
Based on the proposed ACIM performance estimation model, we obtain four objective functions $f_{\text{SNR}}$, $f_{\text{T}}$, $f_{\text{E}}$, $f_{\text{A}}$. Equation~\ref{equ:throughput},~\ref{equ:energy},~\ref{equ:area} clearly demonstrates $f_{\text{T}}$, $f_{\text{E}}$, $f_{\text{A}}$, respectively. The $f_{\text{SNR}}$ can be obtained by simplifying Equations~\ref{equ:SNRTotal}-\ref{equ:SQNR}. The simplified $f_{\text{SNR}}$ is depicted in Equation~\ref{finalSNR}, where $k_{3}$ and $k_{4}$ are constant coefficients related to the data distribution, $C_{\mathrm{o}}$ is the compute capacitor. 

\begin{equation}
    \label{finalSNR}
    \text{SNR}_{\text{(dB)}}=6 B_{\text{ADC}}-10 \log 10 \frac{H}{L}-10 \log 10 \frac{k_{3}}{C_{\mathrm{o}}}+k_{4}
\end{equation}

Based on the previous analysis, the multi-objective optimization problem of ACIM performance can be formulated as Equation~\ref{finalobjetct}. The negative sign in front of $f_{\text{SNR}}$ and $f_{\text{T}}$ means that it is required to solve for the maximum value. The constraint $H - L \geq 0$ guarantees that local array size $L$ can not be larger than the array height $H$ and $\frac{H}{L} - 2^{B_\text{ADC}}$ indicate that the ADC precision is limited by the available capacitors The constraint $ H \cdot W = Arraysize$ guarantees the final array size is exactly equal to the user-defined array size. Finally, a classic NSGA-II algotithm~\cite{pereiraReviewMultiobjectiveOptimization2022} is performed and a high-quality Pareto-frontier set can be obtained efficiently. 

\begin{equation}
    \label{finalobjetct}
    \vspace{-0.3cm}
    \begin{array}{c}
        \min_{x} \quad F(H,W,L,B_{\text {ADC}}) =\left[-f_{\text{SNR}}, -f_{\text{T}}, f_{\text{E}}, f_{\text{A}} \right] \\
        \text { s.t. } \frac{H}{L} - 2^{B_\text{ADC}} \geq 0 \\ 
        H - L \geq 0 \\
        H \cdot W = Arraysize
    \end{array}
\end{equation}

\subsection{Template-based Hierechical Placer and Router}

After obtaining the user-distilled Pareto-frontier set, the template-based ACIM netlist generator generates netlists for each solution in the user-distilled Pareto-frontier set. Due to the page limit, we omit the details on the netlist generator, which follows a straightforward engineering process. Then, EasyACIM performs template-based hierarchical placement and routing for the netlists to generate the final layouts. 
\begin{figure}[b]
    \centering
    \vspace{-0.5cm}
    \includegraphics[width=0.33\textwidth]{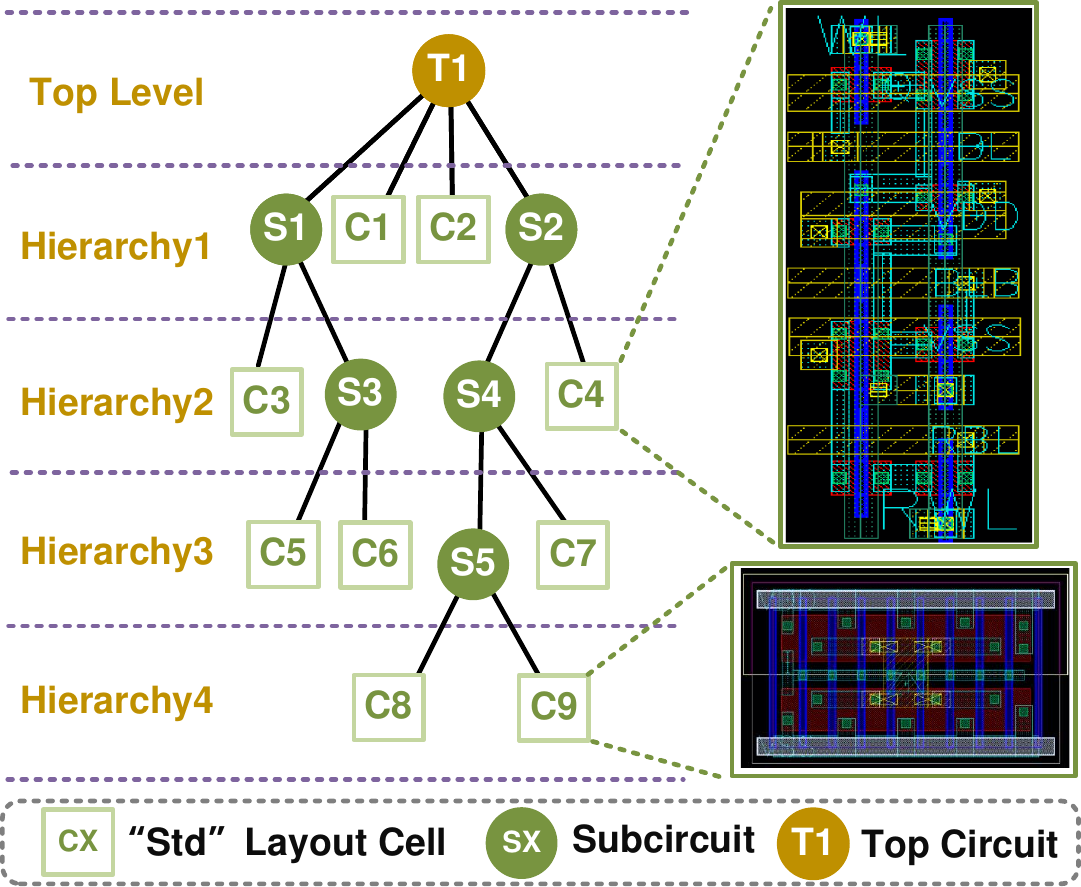}
    \caption{Strategy of template-based hierarchical placer and router.}
    \label{TemplatePR}
\end{figure}

The fundamental of the proposed placer and router is the classic grid-based algorithm~\cite{zhuHierarchicalAnalogMixedSignal2023,zhangSAGERouteSynergisticAnalog,zhangSAGERoute2.0SynergisticAnalog}. However, the fully automated layouts often fail to meet strict design requirements. Therefore, EasyACIM extends the classic algorithm to support manually designed cells in the layout automation framework shooting for better performance. EasyACIM leverages the hierarchical framework to facilitate this extension. Figure~\ref{TemplatePR} depicts the strategy of template-based hierarchical placer and router. The "Std" layout cell indicates either a real standard cell or PCell in PDK, or a manually designed cell. In each hierarchy, the placement and routing inside the "Std" layout cell or subcircuit will be kept, only the inter-connection routing and over-cell placement are conducted. For example, in Hierarchy 1 only the interconnection among C1, C2, S1, and S2 will be routed and these "Std" layout cells or subcircuits will be placed as a whole. Finally, following a bottom-up strategy, the final layout can be generated with manual-designed cells.

\section{Experimental Results}
\label{sec:Results}

\begin{sloppypar}
We perform experiments on a Linux server with an Intel Xeon Gold 6230 CPU @ 2.10GHz. EasyACIM is implemented on the TSMC28 PDK with 1bx1b computation. The agile design exploration for a particular array size can be finished in 30 minutes. The layout generation for a particular solution in the Pareto-frontier set can be done in a few minutes thanks to the customized cell library and pre-defined routing tracks for critical nets including power nets and SAR logic control nets. 
\end{sloppypar}

\begin{sloppypar}
The main differences between EasyACIM and other design flows are shown in Table~\ref{tab:comparison}. Compared to the traditional flow, EasyACIM can dramatically accelerate the design cycle and generate design layouts automatically. In contrast to the AutoDCIM~\cite{chenAutoDCIMAutomatedDigital}, EasyACIM automatically determines the design parameters (e.g. H, W, L, $B_\text{ADC}$) and performs agile design space exploration to uncover the Pareto frontier, while AutoDCIM~\cite{chenAutoDCIMAutomatedDigital} only takes the user-defined design parameters and conducts design space exploration without any optimization. 
\end{sloppypar}

\begin{figure}[tb]
    \centering
    \includegraphics[width=0.3\textwidth]{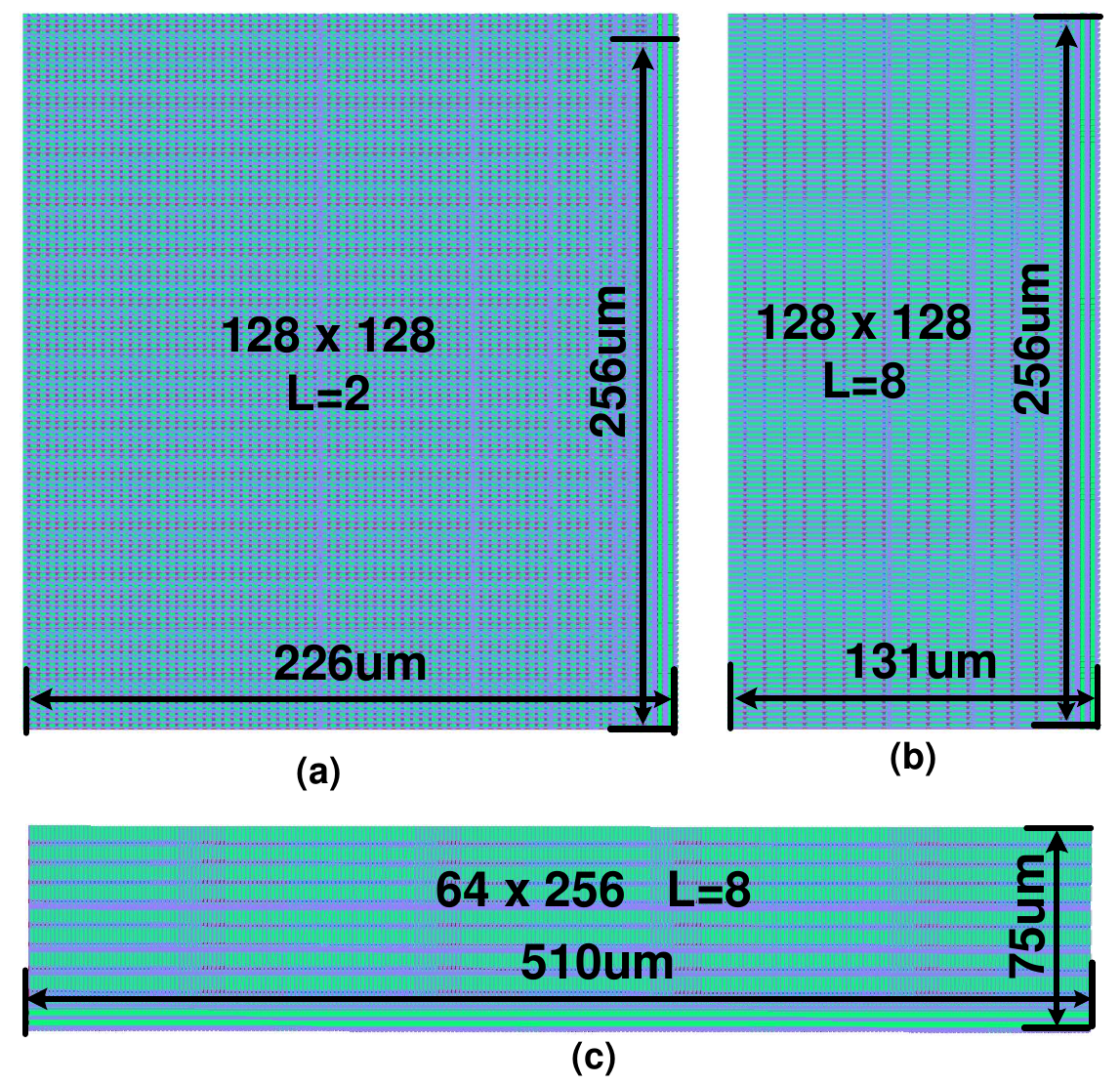}
    \caption{The layouts of 16kb ACIM with various design specifications.}
    \label{LAyout}
    \vspace{-0.6cm}
\end{figure}

\begin{sloppypar}
Figure~\ref{LAyout} demonstrates the final layout results of a 16kb ACIM with 3-bit ADC precision. Figure~\ref{LAyout}\textcolor{ACMPuple}{(a)} shows the situation where $H$=128, $L$=2 shooting for high throughput (3.277TOPS) but at the expense of area (4504$\text{F}^2$/bit). Figure~\ref{LAyout}\textcolor{ACMPuple}{(b)} depicts a design with a more balanced performance (throughput=0.813TOPS,area=2610$\text{F}^2$/bit). Compared to Figure~\ref{LAyout}\textcolor{ACMPuple}{(b)}, Figure~\ref{LAyout}\textcolor{ACMPuple}{(c)} achieved higher SNR and the same throughput at the expense of area (area=2977$\text{F}^2$/bit). 
\end{sloppypar}

\begin{figure*}[tb]
    \centering
    \includegraphics[width=0.9\textwidth]{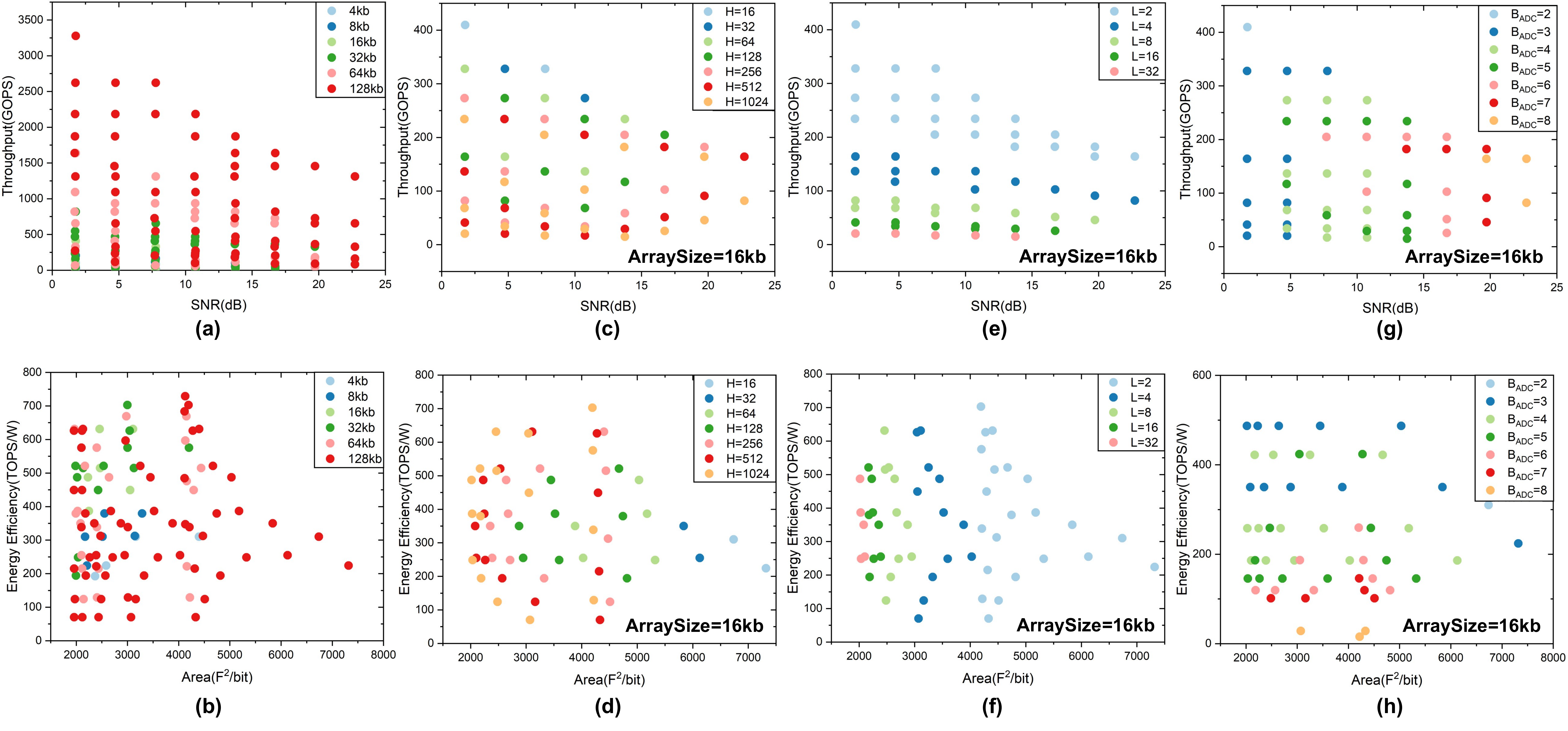}
    \caption{Design Space of EasyACIM: (a) (b) Design space categorized by array size; (c) (d) Design space categorized by $H$ with 16kb array size; (e) (f) Design space categorized by $L$ with 16kb array size; (g) (h) Design space categorized by $B_\text{ADC}$ with 16kb array size.}
    \label{DesignSpace}
    \vspace{-0.3cm}
\end{figure*}

\begin{sloppypar}
    A holistic analysis of EasyACIM design space is shown in Figure~\ref{DesignSpace}. During the design space exploration, $B_\text{ADC}$ is set within 8 bits and $L$ is limited to between 2 and 32 to avoid extreme results.  Figure~\ref{DesignSpace}\textcolor{ACMPuple}{(a)(b)} shows the overall design space of EasyACIM. Figure~\ref{DesignSpace}\textcolor{ACMPuple}{(c)(d)}, Figure~\ref{DesignSpace}\textcolor{ACMPuple}{(e)(f)}, and Figure~\ref{DesignSpace}\textcolor{ACMPuple}{(g)(h)} illustrate the impact of different parameters $H$, $L$, $B_\text{ADC}$ on the design space with a given array size. In Figure~\ref{DesignSpace}\textcolor{ACMPuple}{(a)(b)} it can be seen that larger arrays present the potential to achieve higher SNR and throughput, while smaller arrays prioritize energy efficiency and area. Figure~\ref{DesignSpace}\textcolor{ACMPuple}{(c)(d)} illustrates that a smaller $H$ can lead to a higher throughput. However, this comes with limitations in SNR and an increase in area overhead. Figure~\ref{DesignSpace}\textcolor{ACMPuple}{(e)(f)} depicts that reducing $L$ leads to higher throughput and an increased upper bound of SNR, but incurs additional area overhead. As shown in Figure~\ref{DesignSpace}\textcolor{ACMPuple}{(g)(h)}, reducing $B_\text{ADC}$ enhances energy efficiency, yet it notably diminishes the SNR as well.
\end{sloppypar}

\begin{figure}[tb]
    \centering
    \includegraphics[width=0.3\textwidth]{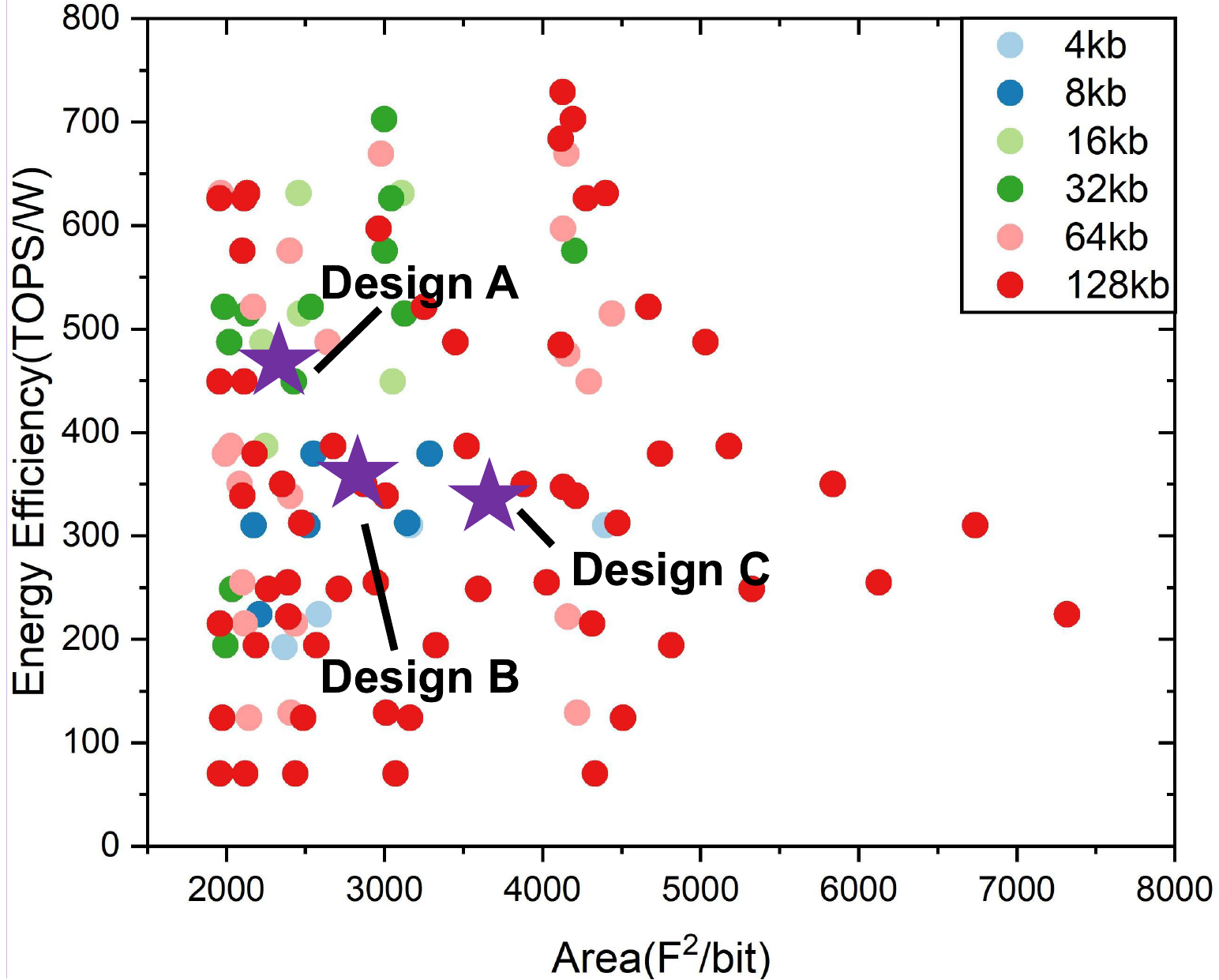}
    \caption{Comparasion between EasyACIM and SOTA ACIMs.}
    \label{SOTA}
\end{figure}

\begin{sloppypar}
    The most common evaluation metrics of ACIM are energy efficiency and area. In Figure~\ref{SOTA}, we compare the design space with the SOTA ACIM designs. Design A~\cite{yaoFullyBitFlexibleComputation2023a}, design B~\cite{yu65nm8TSRAM2022}, design C~\cite{dong15351TOPS3722020} are SOTA ACIMs from JSSC/ISSCC in recent years. The Pareto frontier based on energy efficiency and area is highlighted with blue dashed lines in Figure~\ref{SOTA}. It can be seen that EasyACIM can generate high-quality ACIM solutions with competitive performance to SOTA ACIMs. The ACIM solutions generated by EasyACIM also have wide design space with energy efficiency ranging from 50TOPS/W to 750TOPS/W and area ranging from 1500$\text{F}^2$/bit to 7500$\text{F}^2$/bit. 
\end{sloppypar}

  % \multirow{2}{*}{Benchmark} & Placement  & Technology  & \multirow{2}{*}{Die Size}  \\

\begin{table}[tb]
  \centering
  \caption{Comparison with Other CIM Design Flow.}
  % \vspace{0.2cm}
  \label{tab:comparison}
  \resizebox{0.43\textwidth}{!}{
  \begin{tabular}{c|ccc}
  \toprule
  Entry & Traditional Flow& AutoDCIM~\cite{chenAutoDCIMAutomatedDigital} & EasyACIM \\
  \midrule
  Design type & Analog or Digital & Digital & Analog \\ 
  Design of layout & Manual & Automatic & Automatic \\
  Design time & 1-2 months & NA & Several hours \\ 

  Design space & Fixed & Unoptimized & Pareto frontier  \\
  \midrule
     Determination of & \multirow{2}{*}{Manual}  & \multirow{2}{*}{User-defined}  & \multirow{2}{*}{Automatic}  \\
  design parameters &  &  & \\

  \bottomrule
  \end{tabular}
  }
  \vspace{-0.3cm}
\end{table}

\section{Conclusion}
\label{sec:Conclusion}

\begin{sloppypar}
In this paper, we propose EasyACIM, the first end-to-end automated ACIM. Based on a novel synthesizable architecture, EasyACIM can be easily implemented in various applications with different requirements. Leveraging the MOGA-based Pareto-frontier explorer and template-based hierarchical layout placer and router, EasyACIM can generate high-quality ACIM solutions with competitive 
performance to SOTA ACIMs and wide design space where the energy efficiency ranges from 50TOPS/W to 750TOPS/W and area ranges from 1500$\text{F}^2$/bit to 7500$\text{F}^2$/bit. Validated in TSMC28, the experimental results demonstrate the robustness and benefits of EasyACIM.
\end{sloppypar}

\section{Acknowledgement}
This work was supported in part by the National Science Foundation of China (Grant No. 62141404, 62125401), the Natural Science Foundation of Beijing, China (Grant No. Z230002), and the 111 project (B18001).

\vspace{-.05in}
{
%\scriptsize
\small
\bibliographystyle{IEEEtran}
\bibliography{./ref/DAC24}

\begin{thebibliography}{10}
\providecommand{\url}[1]{#1}
\csname url@samestyle\endcsname
\providecommand{\newblock}{\relax}
\providecommand{\bibinfo}[2]{#2}
\providecommand{\BIBentrySTDinterwordspacing}{\spaceskip=0pt\relax}
\providecommand{\BIBentryALTinterwordstretchfactor}{4}
\providecommand{\BIBentryALTinterwordspacing}{\spaceskip=\fontdimen2\font plus
\BIBentryALTinterwordstretchfactor\fontdimen3\font minus \fontdimen4\font\relax}
\providecommand{\BIBforeignlanguage}[2]{{%
\expandafter\ifx\csname l@#1\endcsname\relax
\typeout{** WARNING: IEEEtran.bst: No hyphenation pattern has been}%
\typeout{** loaded for the language `#1'. Using the pattern for}%
\typeout{** the default language instead.}%
\else
\language=\csname l@#1\endcsname
\fi
#2}}
\providecommand{\BIBdecl}{\relax}
\BIBdecl

\bibitem{vermaInMemoryComputingAdvances2019}
N.~Verma \emph{et~al.}, ``In-{{Memory Computing}}: {{Advances}} and {{Prospects}},'' \emph{IEEE Solid-State Circuits Magazine}, vol.~11, pp. 43--55, 2019.

\bibitem{jhangChallengesTrendsSRAMBased2021}
C.-J. Jhang \emph{et~al.}, ``Challenges and {{Trends}} of {{SRAM-Based Computing-In-Memory}} for {{AI Edge Devices}},'' \emph{TCAS-I}, vol.~68, pp. 1773--1786, 2021.

\bibitem{cheon2941TOPSChargeDomain10T2023}
S.~Cheon \emph{et~al.}, ``A 2941-{{TOPS}}/{{W Charge-Domain 10T SRAM Compute-in-Memory}} for {{Ternary Neural Network}},'' \emph{TCAS-I}, vol.~70, pp. 2085--2097, 2023.

\bibitem{yaoFullyBitFlexibleComputation2023a}
C.-Y. Yao \emph{et~al.}, ``A {{Fully Bit-Flexible Computation}} in {{Memory Macro Using Multi-Functional Computing Bit Cell}} and {{Embedded Input Sparsity Sensing}},'' \emph{JSSC}, vol.~58, pp. 1487--1495, 2023.

\bibitem{yu65nm8TSRAM2022}
C.~Yu \emph{et~al.}, ``A 65-nm {{8T SRAM Compute-in-Memory Macro With Column ADCs}} for {{Processing Neural Networks}},'' \emph{JSSC}, vol.~57, pp. 3466--3476, 2022.

\bibitem{zhangHelloEdgeKeyword2018a}
A.~Biswas \emph{et~al.}, ``An area-efficient 6t-sram based compute-in-memory architecture with reconfigurable sar adcs for energy-efficient deep neural networks in edge ml applications,'' in \emph{CICC}, 2022, pp. 1--2.

\bibitem{yan041MbMm272022}
B.~Yan \emph{et~al.}, ``A 1.041-{{Mb}}/mm {\textsuperscript{2}} 27.38-{{TOPS}}/{{W Signed-INT8 Dynamic-Logic-Based ADC-less SRAM Compute-in-Memory Macro}} in 28nm with {{Reconfigurable Bitwise Operation}} for {{AI}} and {{Embedded Applications}},'' in \emph{ISSCC}.\hskip 1em plus 0.5em minus 0.4em\relax {IEEE}, 2022, pp. 188--190.

\bibitem{dong15351TOPS3722020}
Q.~Dong \emph{et~al.}, ``15.3 {{A 351TOPS}}/{{W}} and 372.{{4GOPS Compute-in-Memory SRAM Macro}} in 7nm {{FinFET CMOS}} for {{Machine-Learning Applications}},'' in \emph{ISSCC}.\hskip 1em plus 0.5em minus 0.4em\relax {IEEE}, 2020.

\bibitem{yuLogicCompatibleEDRAMComputeInMemory2021}
C.~Yu \emph{et~al.}, ``A {{Logic-Compatible eDRAM Compute-In-Memory With Embedded ADCs}} for {{Processing Neural Networks}},'' \emph{TCAS-I}, vol.~68, pp. 667--679, 2021.

\bibitem{ali65Nm462023}
M.~Ali \emph{et~al.}, ``A 65 nm 1.4-6.7 {{TOPS}}/{{W Adaptive-SNR Sparsity-Aware CIM Core}} with {{Load Balancing Support}} for {{DL}} workloads,'' in \emph{CICC}.\hskip 1em plus 0.5em minus 0.4em\relax {IEEE}, 2023, pp. 1--2.

\bibitem{chenNeuroSimCircuitLevelMacro2018}
P.-Y. Chen \emph{et~al.}, ``{{NeuroSim}}: {{A Circuit-Level Macro Model}} for {{Benchmarking Neuro-Inspired Architectures}} in {{Online Learning}},'' \emph{TCAD}, vol.~37, pp. 3067--3080, 2018.

\bibitem{sunAnalogDigitalInmemory}
J.~Sun \emph{et~al.}, ``Analog or {{Digital In-memory Computing}}? {{Benchmarking}} through {{Quantitative Modeling}},'' \emph{ICCAD}, 2023.

\bibitem{gonugondlaFundamentalLimitsPrecision2020}
S.~K. Gonugondla \emph{et~al.}, ``Fundamental limits on the precision of in-memory architectures,'' in \emph{ICCAD}.\hskip 1em plus 0.5em minus 0.4em\relax {ACM}, 2020, pp. 1--9.

\bibitem{gonugondlaFundamentalLimitsEnergyDelayAccuracy2022}
S.~K. Gonugondla \emph{et~al.}, ``Fundamental {{Limits}} on {{Energy-Delay-Accuracy}} of {{In-Memory Architectures}} in {{Inference Applications}},'' \emph{TCAD}, vol.~41, pp. 3188--3201, 2022.

\bibitem{kamineniMemGenOpenSourceFramework2021}
S.~Kamineni \emph{et~al.}, ``{{MemGen}}: {{An Open-Source Framework}} for {{Autonomous Generation}} of {{Memory Macros}},'' in \emph{CICC}.\hskip 1em plus 0.5em minus 0.4em\relax {IEEE}, 2021, pp. 1--2.

\bibitem{liuOpenSAROpenSource2021}
M.~Liu \emph{et~al.}, ``{{OpenSAR}}: {{An Open Source Automated End-to-end SAR ADC Compiler}},'' in \emph{ICCAD}.\hskip 1em plus 0.5em minus 0.4em\relax {IEEE}, 2021, pp. 1--9.

\bibitem{chenAutoDCIMAutomatedDigital}
J.~Chen \emph{et~al.}, ``Autodcim: An automated digital cim compiler,'' in \emph{DAC}, 2023, pp. 1--6.

\bibitem{kangMultiFunctionalInMemoryInference2018}
M.~Kang \emph{et~al.}, ``A {{Multi-Functional In-Memory Inference Processor Using}} a {{Standard 6T SRAM Array}},'' \emph{JSSC}, vol.~53, pp. 642--655, 2018.

\bibitem{yinXNORSRAMInMemoryComputing2020}
Z.~Jiang \emph{et~al.}, ``Xnor-sram: In-memory computing sram macro for binary/ternary deep neural networks,'' in \emph{2018 IEEE Symposium on VLSIT}, 2018, pp. 173--174.

\bibitem{rastegariXNORNetImageNetClassification2016}
A.~Biswas and A.~P. Chandrakasan, ``Conv-ram: An energy-efficient sram with embedded convolution computation for low-power cnn-based machine learning applications,'' in \emph{ISSCC}, 2018, pp. 488--490.

\bibitem{pereiraReviewMultiobjectiveOptimization2022}
J.~L.~J. Pereira \emph{et~al.}, ``A {{Review}} of {{Multi-objective Optimization}}: {{Methods}} and {{Algorithms}} in {{Mechanical Engineering Problems}},'' \emph{Archives of Computational Methods in Engineering}, vol.~29, pp. 2285--2308, 2022.

\bibitem{ngatchouParetoMultiObjective2005}
P.~Ngatchou \emph{et~al.}, ``Pareto {{Multi Objective Optimization}},'' in \emph{{{Intelligent Systems Application}} to {{Power Systems}}}.\hskip 1em plus 0.5em minus 0.4em\relax {IEEE}, 2005, pp. 84--91.

\bibitem{kunalINVITEDALIGNOpenSourcea}
K.~Kunal \emph{et~al.}, ``Invited: Align – open-source analog layout automation from the ground up,'' \emph{DAC}, pp. 1--4, 2019.

\bibitem{chenMAGICALOpenSource2021}
H.~Chen \emph{et~al.}, ``{{MAGICAL}}: {{An Open- Source Fully Automated Analog IC Layout System}} from {{Netlist}} to {{GDSII}},'' \emph{IEEE Design \& Test}, vol.~38, pp. 19--26, 2021.

\bibitem{zhangSAGERouteSynergisticAnalog}
H.~Zhang \emph{et~al.}, ``Sageroute: Synergistic analog routing considering geometric and electrical constraints with manual design compatibility,'' in \emph{DATE}, 2023, pp. 1--6.

\bibitem{zhangSAGERoute2.0SynergisticAnalog}
H.~Zhang \emph{et~al.}, ``Sageroute2.0: Hierarchical analog and mixed signal routing considering versatile routing scenarios,'' in \emph{DATE}, 2024, pp. 1--6.

\bibitem{zhuHierarchicalAnalogMixedSignal2023}
K.~Zhu \emph{et~al.}, ``Hierarchical {{Analog}} and {{Mixed-Signal Circuit Placement Considering System Signal Flow}},'' \emph{TCAD}, vol.~42, pp. 2689--2702, 2023.

\bibitem{tripathiMismatchCharacterizationSmall2014}
V.~Tripathi and B.~Murmann, ``Mismatch {{Characterization}} of {{Small Metal Fringe Capacitors}},'' \emph{TCAS-I}, vol.~61, pp. 2236--2242, 2014.

\bibitem{EnergyADC}
B.~Murmann, ``Mixed-signal computing for deep neural network inference,'' \emph{TVLSI}, vol.~29, pp. 3--13, 2021.

\end{thebibliography}
}

\end{document}